# Electron localizations in double concentric quantum ring

I. Filikhin, S.G. Matinyan, and B. Vlahovic

Department of Physics, North Carolina Central University, 1801 Fayetteville Street, Durham, NC 27707, USA

**Abstract**

We investigate the electron localization in double concentric quantum rings (DCQRs) when a perpendicular magnetic field is applied. In weakly coupled DCQRs, the situation can occur when the single electron energy levels associated with different rings may be crossed. To avoid degeneracy, the anti-crossing of these levels has a place. We show that in this DCQR the electron spatial transition between the rings occurs due to the electron level anti-crossing. The anti-crossing of the levels with different radial quantum numbers provides the conditions for electron tunneling between rings. To study electronic structure of the semiconductor DCQR, the single sub-band effective mass approach with energy dependence was used. Results of numerical simulation for the electron transition are presented for DCQRs of geometry related to one fabricated in experiment.

## Introduction

Quantum Rings (QR) are remarkable meso- and nanostructures due to their non-simply connected topology and attracted much attention last decade. This interest supported essentially by the progress in the fabrication of the structures with wide range of geometries including single and double rings [1-5]. This interest roused tremendously in the connection to the problem of the persistent current in mesoscopic rings [6].Transition from meso - to nano -scale makes more favorable the coherence conditions and permit to reduce the problem to the few or even to single electron.

Application of the transverse magnetic field $B$ leads to the novel effects: Whereas the quantum dots (QDs) of the corresponding shape (circular for two dimensional (2D), cylindrical or spherical for 3D) has degeneracy in the radial $n$ and orbital $l$ quantum numbers, QR due to the double connectedness in the absence of the magnetic field $B$ has degeneracy only in $l$, and the nonzero $B$ lifts the degeneracy in $l$, thus making possible the energy level crossing at some value of $B$ [7], potentially providing the single electron transition from one state to the another.

Use the configurations with double concentric QR (DCQR) reveals a new pattern: one can observe the transition between different rings in analogy with atomic phenomena. For the DCQR, the 3D treatment is especially important when one includes the inter ring coupling due to the tunneling. The dependence on the geometries of the rings (size, shape and etc.) [7, 8] becomes essential.

In the light of above mentioned, it is not surprising that numerous papers were devoted in the recent years to the different aspects of DCQR [9-12].

In the present paper, we visualize interesting features occurring in DCQR composed of GaAs in an $Al_{0.70}Ga_{0.30}As$ substrate [4] and the electron transition between rings under influence of the transverse magnetic field B. Therefore, we concentrate here, in contrast with some previous related papers, to the electron spatial transition between inner and outer rings of DCQR which is accompanied by energy levels transition with different radial quantum numbers $n$. The present work is close, in essence, to Ref. [12] where the effect of a magnetic field on the energy levels of electron and holes for cylindrical shaped DCQR was determined for fixed size and for radial quantum number $n=1,2$, with orbital quantum number $|l|$ changing from 1 to 4.

We will see that the spatial transition of electrons in DCQR between rings occur due to the level anti-crossing providing the conditions for tunneling between rings. In this study we use more realistic confining potentials including ones motivated by the experimental fabrication of QRs [4]. Due to the small sizes of the considering DCQRs, we use approximation in which the non-parabolicity of the conduction band is taken into account. That results a shift energy levels and increases the electron effective mass from the bulk value. Proposed model reproduces the observed PL spectra [4] including levels with different radial quantum numbers. We show principal possibility for the electron trapping in inner ring (or dot) of DCQR which may have application in quantum computing.

## Model

The GaAs DCQRs rings, embedded into the $Al_{0.70}Ga_{0.30}As$ substrate, are considered. We use the single sub-band approach, what is justified due to the relatively large band gap of GaAs. The problem can be expressed by the following Schrodinger equation

$$\left(\hat{H}_{kp} + V_c(\mathbf{r})\right)\Psi(\mathbf{r}) = E\Psi(\mathbf{r}). \tag{1}$$

Here $\hat{H}_{kp}$ is the single band **kp**-Hamiltonian operator $\hat{H}_{kp} = -\nabla \frac{\hbar^2}{2m(r,E)^*}\nabla$, $m^*(\mathbf{r},E)$ is the electron effective mass, and $V_c(\mathbf{r})$ is the band gap potential, $V_c(\mathbf{r})=0$ inside the QR and is equal to $E_c$ outside the QR, where $E_c$ is defined by the conduction band offset for the bulk. The Ben-Daniel-Duke boundary conditions are used on interface of the material of QR and substrate. This consideration was restricted by the electron and heavy hole carriers, and the Coulomb interaction was excluded.

In order to account for the non-parabolic effect, the energy dependence of the carrier effective mass is introduced: $m^*/m_0 = f(E,\mathbf{r})$, where $m_0$ is the free electron mass, and $f(E,\mathbf{r})$ is a function of confinement energy [8,13-15]. This function is described by Kane formula [16]. We used the linear dependence of the electron effective mass on energy. The effective mass in QR varies between the bulk values for effective mass of the QR and substrate materials and the energy is rearranged by the quantum well depth. The Schrödinger equation (1) with the energy dependence of effective mass is solved by the iteration procedure. For each step of the iterations the equation (1) is reduced to Schrödinger equation with the effective mass of the current step which does not depend on energy. Obtained eigenvalue problem is solved numerically by the finite element method. After that, a new value for effective mass is taken by using $f(E,\mathbf{r})$ and procedure is repeated. The convergence of the effective mass during the procedure has a place after 3-5 steps.

For each step of the procedure the Schrödinger equation (1) is written in cylindrical coordinates, with constant magnetic field in the z direction ($\mathbf{B} = B\hat{z}$), as follows:

$$-\frac{\hbar^2}{2}\left(\frac{\partial}{\partial\rho}\left(\frac{1}{m^*}\frac{\partial\Phi_{n,l}}{\partial\rho}\right) + \frac{1}{m^*\rho}\frac{\partial\Phi_{n,l}}{\partial\rho} - \frac{l^2}{m^*\rho^2}\Phi_{n,l}\right) + \frac{\hbar l\omega_c}{2}\Phi_{n,l} + \frac{m^*\omega_c^2\rho^2}{8}\Phi_{n,l} + [V_c(\rho,z)-E]\Phi_{n,l} -$$

$$-\frac{\hbar^2}{2m^*}\frac{\partial^2 \Phi_{n,l}}{\partial z^2} = 0, \qquad (2)$$

$\Psi_{n,l}(\rho,z,\varphi) = \Phi_{n,l}(\rho,z)e^{il\varphi}$, where $n = 1,2,3...$ are radial and $l = \pm 0, \pm 1, \pm 2,...$ are orbital quantum numbers. $\omega_c = |e|B/m^*$ is the cyclotron frequency. First magnetic field term in (2) is orbital Zeeman term, the second - diamagnetic term. The electron spin Zeeman effect has been ignored here since it is small.

The values $m^* = 0.067\, m_0$ and $0.093\, m_0$ is used for the bulk values of the electron effective masses in the materials of DCQR and substrate respectively. The effective mass of DCQR differs from the bulk value, and this difference is defined by energy of the electron in the considered state. The confinement potential $V_c(\mathbf{r})$ is zero in the rings and 0.262 eV in the substrate [4]. The contribution of strain was ignored in this paper because the lattice mismatch between the rings and the substrate is small. For the effective mass of heavy hole we used the value of $0.51\, m_0$ in GaAs and $0.57\, m_0$ in Al$_{0.70}$Ga$_{0.30}$As. The confinement potential has the value of 0.195 eV [4].

There is a problem of notation for states for DCQR. If we consider single QR (SQR) then for each value of the orbital quantum number $|l| = 0,1,2,...$ in Eq. (2) we can definite radial quantum number $n = 1,2,3,....$ corresponding to the numbers of the eigenvalues of the problem (2) in order of increasing. One can organize the spectrum by sub-bands defined by different $n$. When we consider the weakly coupled DCQR, in contrast of SQR, the number of these sub-bands is doubled due to known splitting the spectrum of double quantum object [17]. Electron in the weakly coupled DCQR can be localized in the inner or outer ring. In principle, in this two rings problem one should introduce a pair of separate sets of quantum numbers $(n_i, l)$ where index $i = 1,2$ denoted the rings where electron is localized. However, it is more convenient, due to the symmetry of the problem, to have one pair $(n,l)$ numbers ascribed to the different rings (inner or outer), in other words, we use a set of quantum numbers $(n,l), p$ where $p$ is dichotomic parameter attributed to the electron localization ("inner" or "outer").

Since we are interested here in the electron transition between rings and, as we will see below, this transition can occur due to the electron levels anti-crossing followed a tunneling, we concentrate on the changing of the quantum numbers $n$. The anti-crossing is accompanied by changing the quantum numbers $n$ and $p$ of the $(n,l), p$ set.

## Electronic properties of the experimentally fabricated DCQR

The GaAs QRs and DQRs rings, embedded into the Al$_{0.3}$Ga$_{0.7}$As substrate, are considered. We use the geometry of the DCQR which are close to one proposed in [4]. The DCQR cross section is presented in Fig. 1a. Since a profile of the quantum dots in [4] was not explicitly given, this geometry slightly differs from original one used in [4]. However our calculations for the electron energy with this geometry lead to similar results as it was obtained in [4]. To compare with results of [4], we neglected the non-parabolic effect, discussed above, and use the effective masses of the carriers, and the confinement potentials, as in [4]: $m^*_{GaAs}/m^*_{AlGaAs} = 0.067/0.093$, $E_c = 262$ meV for the electron, and $m^*_{GaAs}/m^*_{AlGaAs} = 0.51/0.57$, $E_c = 195$ meV the for heavy hole.

An interesting problem related to the GaAs self-assembled structures deals with the increase of the electron effective mass of QD, QR and DQR, respectively. The non-parabolic effect leads to a change in the effective mass of the carriers in quantum nanosize objects [1,13]. The initial GaAs QDs are quite large in size [4]; therefore for the QD this effect is minimal. The effective electron mass of the QD is practically equal to a bulk mass of the GaAs. It should be noted that the fabrication of DCQRs is accompanied by decreasing of the object size in one or two directions. Taking the energy dependence of the electron effective mass into account, we calculated the effective masses of the DCQR. The results of the calculations are shown in Fig. 1b, where a simple energy dependence of the effective mass (function $f(E,\mathbf{r})$ in Eq.(1)) is represented as a linear function connecting the points corresponding to the bulk values of the effective mass in GaAs and AlGaAs materials, respectively. We obtained for the electron effective mass of the ground state the value of $0.074\,m_0$, which is slightly larger than the bulk value of $0.067\,m_0$. For the excited states, the effective mass will increase with respect to the bulk value of the AlGaAs substrate. In Fig. 1c the few energy peaks of the optical transmission spectrum are shown along the experimental data [4] and calculated results. One can see that our calculated levels are shifted relative the results of [4] due the non-parabolic effect. The effect can be neglected for first two levels, but for higher levels it becomes to be important. The experimental peaks are well reproduced by our calculations.

## Electron transer between rings of DCQR in magnetic field

Electron transfer in the DCQR considered is induced by external factor like a magnetic or electric fields. Probability for this transfer strongly depends on the geometry of DCQR. The geometry has to allow the existing the weakly coupled electron states. To explain it, we note that DCQR can be described as a system of double quantum well. It means that there is duplication of two sub-bands of energy spectrum [17,18] relative the one for single quantum object. In the case non interacting wells (no electron tunneling between wells) the each sub-band is related with left or right quantum well. The wave function of the electron is localized in the left or right quantum well. When the tunneling is possible (strong coupling state of the system), the wave function is spread out over whole volume of the system. In a magnetic field, it is allowed an intermediate situation (weak coupled states) when the tunneling is possible due to anti-crossing of the levels.

  Strongly localized states are existed in the DCQR with the geometry motivated by the fabricated DCQR in Ref. [4]. The wave functions of the two $s$-states of the single electron with $n=1,2$ are shown in Fig. 2, where the electron state $n=1$ is localized in outer ring, and the electron state $n=2$ is localized in inner ring. Moreover all states of the sub-bands with $n=1,2$, and $|l|=1,2,3,...$, are well localized in the DCQR. The electron localization is in outer ring for $n=1$, $|l|=0,1,2...$, and in inner ring for $n=2$, $|l|=0,1,2...$. It is shown in Fig. 3 for several low-lying electron levels. The difference between spectra of the two sub-bands presented Fig. 3 can be explained by competition of two terms of the Hamiltonian of Eq. (2) and geometry factor. The first term includes first derivative of wave function over $\rho$ in kinetic energy; the second is the centrifugal term. For $|l|\neq 0$ the centrifugal force pushes the electron into outer ring. One can see that the density of the levels is higher in the outer ring. Obviously, the geometry plays a role also. In particular, one can regulate density of levels of the rings by changing a ratio of the lateral sizes of the rings.

  Summarizing, one can say that for $B=0$ the well separated states are only the states $(1,l), p$ and $(2,l), p$. Thus, used notation is proper only for these states. The wave functions of the rest states $(n>2,l)$ are distributed between inner and outer rings. These states are strongly coupled states.

  In Ref. [12] the $(2,l), p$ and $(1,l), p$ states are denoted as the L and H states, respectively. We have to note that difference of the both descriptions is in that the notation [12] does not describe the position change of the electron in the states $(2,l), p$ or $(1,l), p$ under increasing magnetic field.

Crossing of electron levels in the magnetic field $B$ are presented in Fig. 4. There are crossings of the levels without electron transfer between the rings. This situation is like when we have crossing levels of two independent rings. There are two crossings when the orbital quantum number is changed due to the Aharonov-Bohm (AB) effect. It occurs at about 0.42 T and 2.5 T. There are two anti-crossings: the first is at 4.8 T, another is at 5.2 T. These anti-crossings are for the states with different $n$; the first are states (1,0) and (2,0) and the second are states (1,-1) and (2,-1). In these anti-crossings the possibility for electron tunneling between rings are realized. In Fig. 5 we show how the root mean square (rms) of the electron radius is changed due to the tunneling at anti-crossing. One can conclude from Fig. 4-5 that the electron transition between rings is only possible when the anti-crossed levels have different radial quantum numbers.

Transformation of the profile of the electron wave function during the process of anti-crossing with increasing $B$ is given in Fig. 6. The electron state $(1,-1)$, outer is considered as "initial" state of an electron ($B=0$). The electron is localized in outer ring. Rms radius is calculated to be $R=39.6$ nm. For $B=5.2$ T, the second state is the tunneling state corresponding to the anti-crossing with the state (0,-1). The wave function is speeded out in both rings with $R=32.7$ nm. The parameter $p$ has no definite value for this state. The "final" state is considered at $B=7$ T. In this state the electron localized in inner ring with $R=17.6$ nm. Consequently connecting these three states of the electron, we come to an electron trapping, when the electron of outer ring ("initial" state) is transferred to the inner ring ("final" state). The transfer process is governed by the magnetic field.

In the case of planar QRs ($H << R$) the relationship between the energy and the magnetic flux $\Phi$ can approximately be described by the following relation for the ideal quantum ring of radius $R$ in a perpendicular magnetic field $B$: $E_p(l) = \hbar^2/(2m^*R_p^2)(l+\Phi/\Phi_0)^2$, where $\Phi = \pi R^2 B$, $\Phi_0 = h/e$, and $p$ means inner or outer ring, $\Phi_0 = 4135.7$ Tnm$^2$. It is clear that this relation leads to the periodic oscillations of the energy with the AB period $\Delta B = \Phi_0/\pi R^2$. Using rms radius as $R$, one can obtain for the inner ring $\Delta B/2 = 2.1$ T, for the outer ring $\Delta B/2 = 0.42$ T. The $R$s are 17.7 nm and 39.6 nm, respectively. The obtained values for $\Delta B/2$ corresponding to the level crossing $(n,0)$ and $(n,-1)$, where $n=2,1$, shown in Fig. 4, about at 2.5 T and 0.42 T, respectively. Thus, this rough estimation qualitatively reproduces the results for the AB period presented in the Fig. 4.

Note that the energy gap between anti-crossed levels, which one can see in Fig. 4, can be explained by the general theory for double interacting quantum well (see [18], for instance). The value of the gap depends on separation distance between the rings, governed by the overlapping wave functions corresponding to the each ring, and their spatial spread which mainly depends on radial quantum number of the states.

## Conclusion

Nanosize quantum rings were studied in the single sub-band approach, taking into account the non-parabolicity of the conduction band due to energy dependence of the electron effective mass. Realistic 3D geometry relevant to the experimental DCQR fabrication, was considered.

We make visible main properties of this weakly coupled DCQD established by several level anti-crossings that occurred for the states with different radial quantum number $n$ ($n=1,2$). In the present paper, in contrast to the many publications on the subject based on the parabolic confinement potentials, we used more realistic potentials including ones motivated by the experimental fabrication of the QRs [4].

We may conclude that the fate of the single electron in DCQRs is governed by the structure of the energy levels with their crossing and anti-crossing and changing with magnetic field. The above described behavior is the result of the nontrivial excitation characteristic of the DCQRs. Effect of the trapping of electron in inner QR of DCQR may be interesting from the point of view of quantum computing.

## Acknowledgements

This work is supported by the NSF (HRD-0833184) and NASA (NNX09AV07A).

**Figures:**

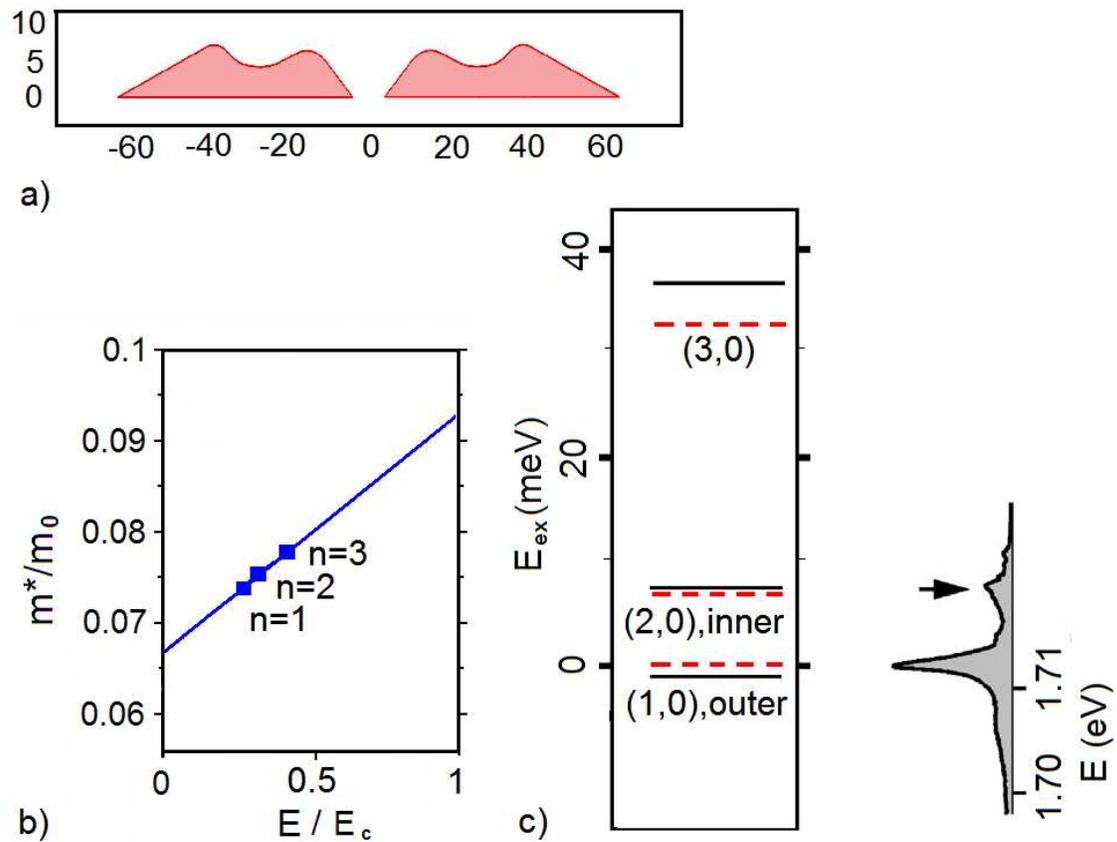

Figure 1. a) Cross section of DCQR. The sizes are given in nm. b) The effective mass and energy of electron in the states $n = 1,2,3$ ($l = 0$) c) The optical transmission spectrum and experimental data from [4]. The solid lines are results of calculation [4]; dashed lines are results of the present work. The arrow shows the (2,0) pick of the PL transition relative the (1,0) exciton energy.

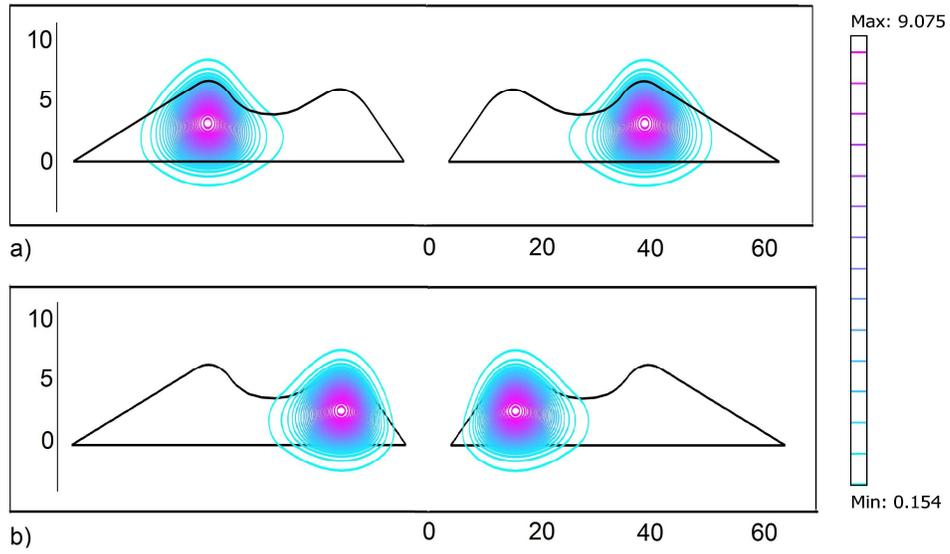

Figure 2. The squares of wave functions for the a) $(1,0)$, outer ($E = 0.072$ eV) and b) $(2,0)$, inner ($E = 0.080$ eV) states are shown by contour plots. The contour of the DCQR cross-section is given. The sizes are in nm.

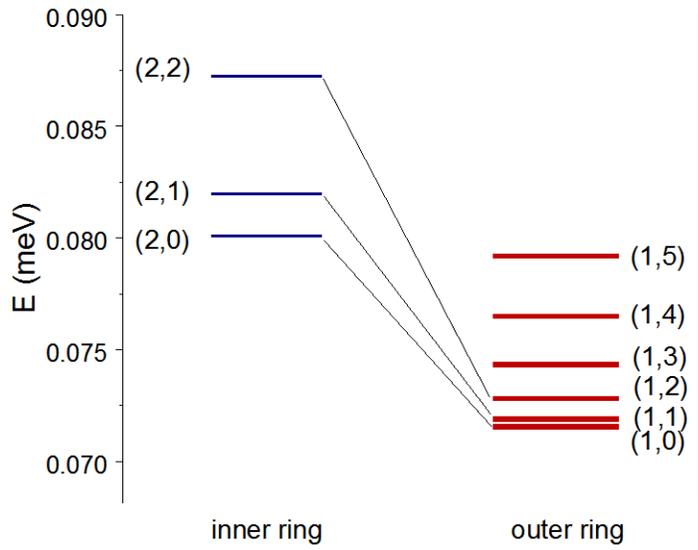

Figure 3. Electron energy $E$ and position of single electron in DCQR for the states with $n = 1,2$, $l = 0,1,2,...5$. The quantum number each state is shown. Fine lines connect the upper and lower members of the "doublets" $((n=1,2),l)$, for $l = 0,1,2$. Energy is measured from bottom edge of conduction bands alignment.

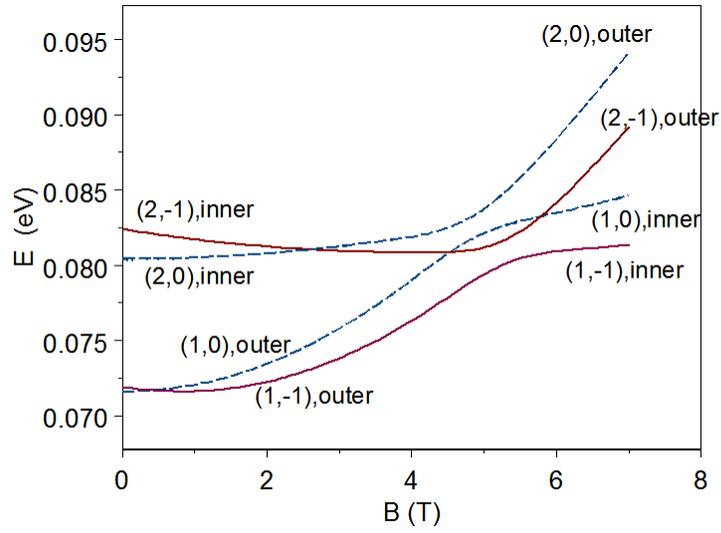

Figure 4. Single electron energies of DCQR as a function of magnetic field magnitude $B$. Notation for the curves: the double dashed (solid) lines mean states with $l=0$ ($l=-1$) with $n=1,2$. The quantum numbers of the states and positions of the electron in DCQR are shown.

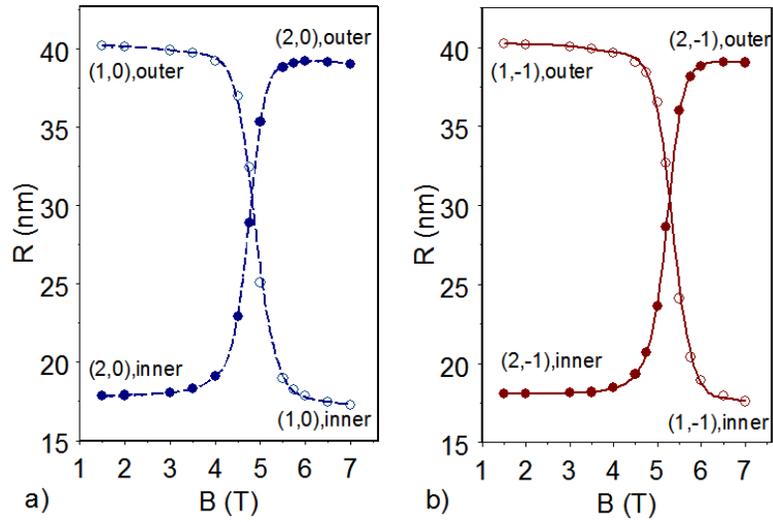

Figure 5. Rms radius of an electron in DCQR as a function of magnetic field for the states a) $((n=1,2), l=0)$ and b) $((n=1,2), l=-1)$ near point of the anti-crossing. The calculated values are shown by solid and open circles. The dashed (solid) line, associated with states of $l=0$ ($l=-1$), fits the calculated points.

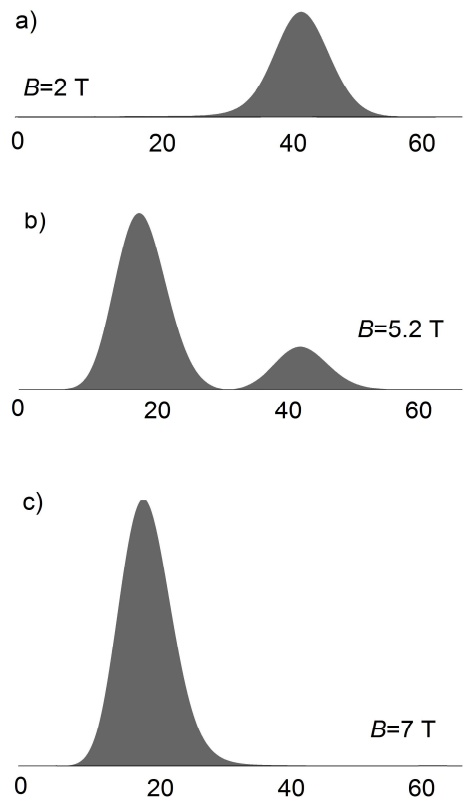

Figure 6. Profiles of the normalized square wave function of electron in the states a) $(1,-1)$,outer; b) $(1,-1)$,n/a and c) $(1,-1)$,inner for different magnetic field $B$. a) is the "initial" state ($B=0$) with $R=39.6$ nm, b) is the state of electron transfer ($B=5.2$ T) with $R=32.7$ nm, c) is the "final" state ($B=7$ T) with $R=17.6$ nm. The radial coordinate $\rho$ is given in nm (see Fig. 1 for the DCQR cross section).